\documentclass[aps,prl,reprint]{revtex4-1}
\usepackage{graphics} 
\usepackage{graphicx} 
\usepackage{dcolumn}
\usepackage{bm}
\usepackage{braket}
\usepackage[english]{babel}
\usepackage{color}
\usepackage{chemist}
\usepackage{amsfonts}
\newcommand{\red}{\textcolor{black}}

\begin{document}
\title{Schwinger boson theory of the $J_1,J_2=J_{3}$ kagome antiferromagnet }
\author{Tristan Lugan$^{1}$, Ludovic D.C. Jaubert$^{2}$, Masafumi Udagawa$^{3}$, Arnaud Ralko$^{1}$}
\affiliation{$^{1}$Institut N\'eel, UPR2940, Universit\'e Grenoble Alpes et CNRS, Grenoble, FR-38042 France\\
$^{2}$Universit\'e de Bordeaux, CNRS, LOMA, UMR 5798, FR-33405 Talence, France \\
$^{3}$Department of Physics, Gakushuin University, Mejiro, Toshima-ku, Tokyo 171-8588, Japan}

\date{\today}

\begin{abstract}
We study the kagome antiferromagnet for quantum spin$-1/2$ with first $J_{1}$, second $J_{2}$ and third $J_{3}$ neighbour exchanges, along the $J_2 = J_{3}\equiv J$ line. We use Schwinger-boson mean-field theory for the precise determination of the phase diagram, and two different rewritings of the Hamiltonian to build an intuition about the origin of the transitions. The spin liquid obtained at $J=0$ remains essentially stable over a large window, up to $J\approx 1/3$, because it is only weakly frustrated by the $J$ term. Then at $J\approx 1/2$, the intermediate $Z_{2}$ spin liquid condenses into a long-range chiral order because of the change of nature of local magnetic fluctuations. As a side benefit, our Hamiltonian rewriting offers an exact solution for the ground state of our model on a Husimi cactus.
\end{abstract}
\maketitle

\begin{figure}[t]
\centering\includegraphics[width=0.5\textwidth]{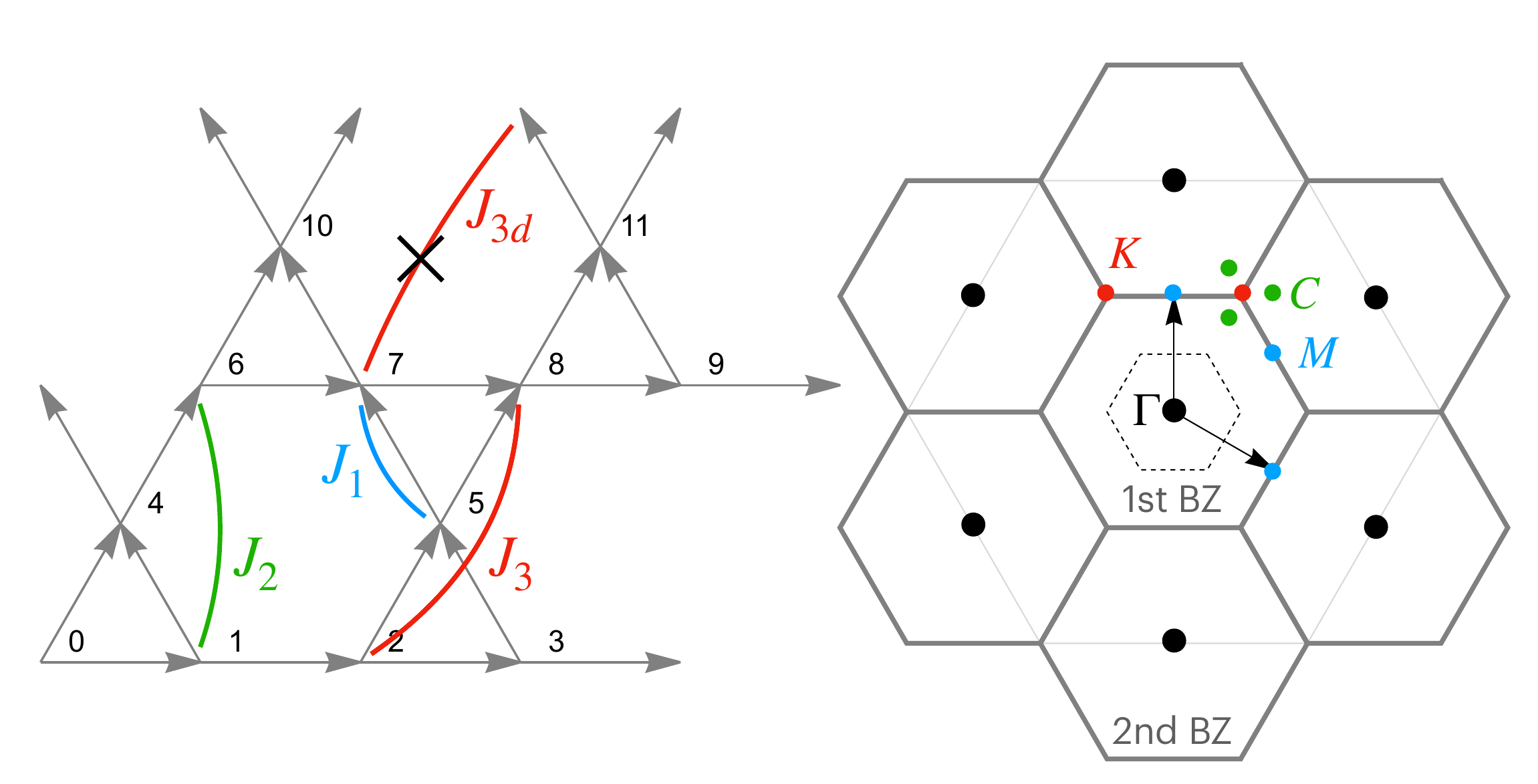}
\caption{
\textit{Left:} The 12-site unit-cell of the kagome lattice used in the Schwinger boson theory with the first $J_{1}$, second $J_{2}$ and third $J_{3},J_{3d}$ neighbour couplings. In this paper we  set the energy scale with $J_{1}=1$ and consider $J_{2}=J_{3}\equiv J$ and $J_{3d}=0$. The arrows show the bond orientations of the $A$ and $B$ parameters for nearest-neighbors. The orientations for the $J_2$ and $J_{3}$ bonds are not displayed.
\textit{Right:} The first Brillouin zone (BZ) of the 12-site unit cell (dashed hexagon) is shown, as well as the first and second BZs of the kagome lattice. Relevant high symmetry points are displayed.
}
\label{fig:1}
\end{figure}

The Heisenberg kagome antiferromagnet (HKA) is a canonical model of frustrated magnetism. There is now a relative consensus that its ground state is a quantum spin liquid (QSL), the nature of which -- gapless or not -- remains, however, hotly debated \cite{yan11a,Iqbal11a,messio2012,Mendels16,Lauchli16a,He17a,Ralko2018}. A remarkable property of the HKA is that its ground state is stable for a finite range of perturbations, such as Dzyaloshinskii-Moriya interactions relevant to Herbertsmithite \cite{Mendels16} or further neighbour exchange $J_{1}-J_{2}-J_{3d}$ \cite{Balents02a,he14a,gong14a,bieri2015,iqbal2015} [Fig.~\ref{fig:1}]. The latter Hamiltonian has been actively studied for 20 years \cite{Balents02a}, when it was rewritten as a plaquette Hamiltonian along the $J_{2}=J_{3d}$ line \cite{Balents02a,Palmer01a}. At finite $J_{2},J_{3d}$ values, the HKA spin liquid evolves into a Kalmeyer-Laughlin chiral spin liquid \cite{he14a,gong14a}, a magnetic analogue of the topological order in the fractional quantum Hall effect \cite{Kalmeyer87a}, and connected to the physics of the Kapellasite material \cite{bieri2015,iqbal2015}. Perturbations beyond the HKA are thus a fertile ground for exotic quantum phenomena. 

In this context, one cannot fail to notice that the kagome structure has two inequivalent types of third neighbour couplings: $J_{3d}$ and $J_{3}$ [Fig.~\ref{fig:1}]. As opposed to its more popular counterpart, the $J_{1}-J_{2}-J_{3}$ Hamiltonian has been largely forsaken, even though there is a priori no reason to favour one model over the other. Indeed, here also along the $J_{2}=J_{3}$ line, this spin Hamiltonian has been recently rewritten as a lattice model of interacting topological charges \cite{mizoguchi2017,mizoguchi2018}. For antiferromagnetic $J$, same-charge quasi-particles counter-intuitively \textit{attract} each other, revealing unconventional magnetic textures where fractionalised excitations become stable in the ground state. These works were, however, classical \cite{mizoguchi2017,mizoguchi2018}. On the quantum front the $J_{3}$ coupling alone has lately attracted some interest \cite{Bernu20a,Grison20a,bose22a}, but as far as we know, the $J_{2}=J_{3}$ line has only been considered in \cite{buessen2016,li2021} using a pseudo-fermion functional renormalization group approach (pf-FRG), fermionic mean-field theory and exact diagonalisation. However, it was not discussed in the context of interacting topological charges.

It is the goal of this paper to present a complementary, bosonic, calculations of the $J_{2}=J_{3}$ kagome phase diagram [Eq.~(\ref{eq:ham})], using an unrestricted algorithm of Schwinger-boson mean-field theory (SBMFT) \cite{Ralko2018, lugan2019}. Within SBMFT, we find that the HKA spin liquid evolves into a $Z_{2}$ spin liquid before forming chiral long-range order. Our results are discussed in the context of two Hamiltonian mappings, building an intuition as to the origin of the observed phase boundaries.\\

\textbf{Model.} We consider a system of $n_{s}$ Heisenberg spins$-1/2$ with $J_{1}=1$ and $J_{2}=J_{3}\equiv J$ [Fig.~\ref{fig:1}].
\begin{equation}
\mathcal{H} = \sum_{\langle i,j\rangle_{1}} \hat\mathbf{S}_{i} \cdot \hat\mathbf{S}_{j}
\, +\, J\left(
\sum_{\langle i,j\rangle_{2}} \hat\mathbf{S}_{i} \cdot \hat\mathbf{S}_{j}
+
\sum_{\langle i,j\rangle_{3}} \hat\mathbf{S}_{i} \cdot \hat\mathbf{S}_{j}
\right)
\label{eq:ham}
\end{equation}
%

\textbf{Method.} We study Hamiltonian~(\ref{eq:ham}) by means of SBMFT which treats on an equal footing magnetically ordered and spin-liquid disordered phases \cite{Auerbach1994}. A spin at site $i$ is decoupled as follows
\begin{eqnarray}
{\bf \hat{S}}_i &=& \frac{1}{2} \sum_{\alpha \beta} \hat{b}^+_{i \alpha} \vec{ \sigma}_{\alpha \beta} \hat{b}_{i \beta}
\label{eq:Sb}
\end{eqnarray}
where $\vec{ \sigma}$ are the Pauli matrices, $\hat{b}^{(+)}$ are bosonic operators, and $\alpha,\beta = \uparrow, \downarrow$ are spin directions along the quantization axis perpendicular to the lattice plane. Let us recall the main lines of the SBMFT. More details can be found in \cite{Auerbach1994,Halimeh2016,schaffer2017,Ralko2018,lugan2019} and references therein. First, the Hilbert space is enlarged by the mapping of Eq.~(\ref{eq:Sb}). For a spin $S$, it is thus necessary to enforce the constraint $\hat{n}_i = \hat{b}^+_{i\uparrow}\hat{b}_{i\uparrow}+\hat{b}^+_{i\downarrow}\hat{b}_{i\downarrow}=2S$ on all sites in order to project the solution back onto the physical space. At the mean field level, this is achieved on average by minimizing the free energy with respect to Lagrange multipliers $\lambda_i$ and introducing two SU(2)-invariant bond operators \cite{Flint2009}; the singlet operator $\hat{A}_{ij} = \frac{1}{2} (\hat{b}_{i\uparrow}\hat{b}_{j\downarrow}-\hat{b}_{i\downarrow}\hat{b}_{j\uparrow})$ and the spinon hopping term $\hat{B}_{ij} = \frac{1}{2} (\hat{b}^+_{i\uparrow}\hat{b}_{j\uparrow}+\hat{b}^+_{i\downarrow}\hat{b}_{j\downarrow})$. \red{The latter is a typical measure of magnetic order (where spinons can hop), while the former is favoured in disordered phases made of singlets.} Performing a mean field decoupling on Eq.~(\ref{eq:ham}), we obtain the SBMFT Hamiltonian
\begin{eqnarray}
\mathcal{H}_{\rm SB} &=& \sum_{i,j} J_{ij} \left[ \hat{B}_{ij}^+ B_{ij}+ \hat{B}_{ij} B_{ij}^*- \hat{A}_{ij}^+ A_{ij} - \hat{A}_{ij} A_{ij}^* \right] \nonumber \\
&-& \sum_{i,j} J_{ij} \left[ | B_{ij} |^2 - | A_{ij} |^2\right] + \sum_i \lambda_i \left[ \hat{n}_i - 2 S \right],
\label{eq:hamsbmft}
\end{eqnarray}
with mean field parameters, $A_{ij} = \langle \phi_0 | \hat{A}_{ij} | \phi_0 \rangle$ and $B_{ij} = \langle \phi_0 | \hat{B}_{ij} | \phi_0 \rangle$, as expectation values in the ground state \red{$|\phi_0 \rangle$ for} each oriented pair of interacting spins ($i\rightarrow j$) [Fig.~\ref{fig:1}]. We define a magnetic unit-cell of $n_u$ sites that contains a total number of $12\, n_u $ complex mean-field parameters. We have tried unit-cells up to $n_u = 36$ and found no noticeable differences with $n_u = 12$, the smallest unit cell compatible with all competitive Ans\"atze considered in this work. In the rest of the paper, we thus consider the ($n_u = 12$) unit cell. Eq.~(\ref{eq:hamsbmft}) is solved numerically in a self consistent way, starting from random mean-field parameters $\{A_{ij},B_{ij}\}$ and searching for the set of Lagrange multipliers $\{\lambda_s\}$ satisfying the boson constraint. This last step is achieved by using a least-square minimization. Since all Ans\"atze encountered in this work are translationally invariant, it is enough to consider one Lagrange multiplier per site in the unit-cell, $\{ \lambda_s\}_{s=0,\cdots,n_u-1}$. Ground state $| \phi_0 \rangle$ is obtained by diagonalization -- using a Cholesky decomposition \cite{Toth2015}-- of $(2 n_u) \times (2 n_u)$ q-dependent Hamiltonians written in the Fourier space on a Brillouin zone of linear size $l$ containing $l \times l$ momenta (thus $n_s = n_u \times l \times l$ kagome sites). A new set of mean-field parameters is then computed by using $| \phi_0 \rangle$, and the same procedure is repeated until convergence is reached up to a desired tolerance on mean-field variables, typically $10^{-11}$. \red{$|\phi_0 \rangle$ corresponds to the $T=0$ boson vacuum whose gap scales like $\sim 1/l$ for an ordered phase; condensation only appears in the thermodynamic limit.} We emphasise here that our solutions are unconstrained \cite{Cabra2011,Zhang2013}, and do not a priori assume particular symmetries. The way we update the set of parameters also allows for a derivative-free formulation of the theory that can treat at once complex mean field solutions. We noticed this approach was more stable than an explicit minimisation of the free energy. As a final comment, in the Schwinger boson language $\langle \hat\mathbf{S}^{2}\rangle = 3S(S+1)/2$. This is why \red{we work with the commonly used} spin value $S = \frac{1}{2} ( \sqrt{3} -1)$ in order to recover the good quantum number $\langle \hat\mathbf{S}^{2}\rangle = 3/4$ of a quantum spin$-1/2$ \cite{messio2012}.\\

\begin{figure}[t]
\includegraphics[width=0.5\textwidth]{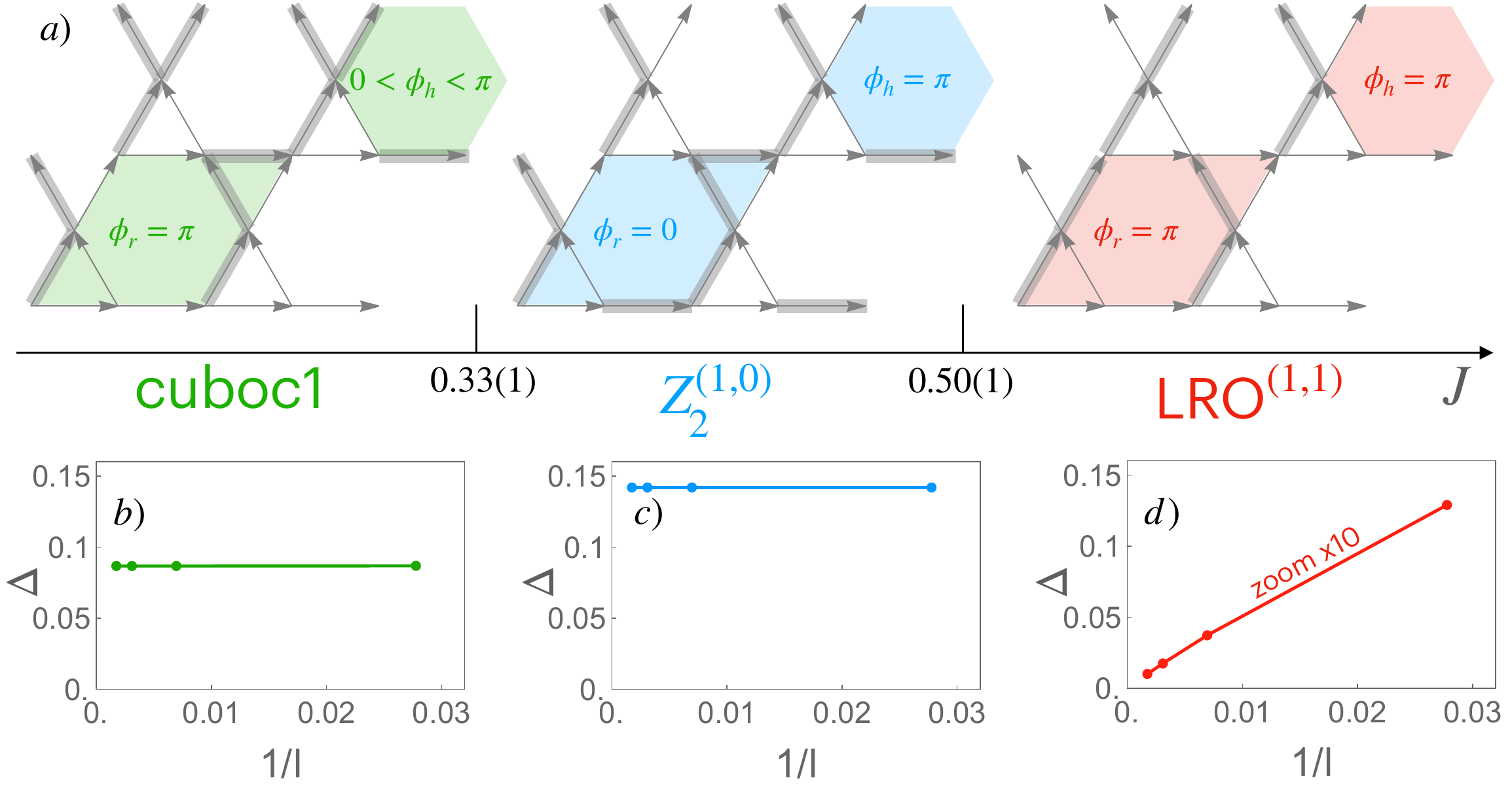}
\caption{
(a) SBMFT phase diagram of the Kagome antiferromagnet along the $J_2=J_{3} = J$ line. Three phases are identified: a  chiral spin liquid (\textit{cuboc1}) \cite{messio2012,Messio2013}, a $Z_2$ spin liquid ($Z_2^{(1,0)}$) originating from the $q=0$ Ansatz, and a magnetically ordered phase (LRO$^{(1,1)}$) coming from the Bose condensation of its spin liquid counterpart, the $Z_2^{(1,1)}$. For each ground state, the sign structure of the nearest neighbours $A_{ij}$ are given, with thin (thick) lines corresponding to positive (negative) values of $A_{ij}$. For all phases, nearest neighbours $|A_{ij}| = \mathcal{A}$. For the \textit{cuboc1}, the arguments of $A_{ij}$ are different between up and down triangles. Phases are also characterised by the flux $\phi_r$ on rhombii and $\phi_h$ on hexagonal Wilson loops.
(b-d) Finite-size scaling of the energy gap $\Delta$ above the ground state for $J=\{0.2, 0.45, 0.6\}$ with $n_{s}=6\, l^{2}$. The y-axis in panel (d) has been multiplied by 10. The spin length in SBMFT is $S=(\sqrt{3}-1)/2$ which gives the good quantum number $\langle \hat\mathbf{S}^{2}\rangle = 3/4$ of a quantum spin$-1/2$ \cite{messio2012}.
\label{fig:2}
}
\end{figure}

\textbf{Observables.} The \red{inelastic} structure factor is a useful tool to visually identify phases, irrespectively of whether they are ordered or not,
\red{
\begin{eqnarray}
S(\mathbf{q},\omega) =\frac{1}{n_s} \sum_{m,n} e^{i \mathbf{q} (\mathbf{r}_m-\mathbf{r}_n)} \int_{\infty}^{\infty} dt e^{- i \omega t} \langle  \hat{\mathbf{S}}_{m}(t) \cdot \hat{\mathbf{S}}_{n} \rangle
\label{eq:Sqw}
\end{eqnarray}
where the sum runs over all $n_s$ sites.} Details about the derivation are given in Ref.~\cite{Halimeh2016}. \red{The equal-time structure factor $S(\mathbf{q})$ is obtained by integrating over all frequencies $\omega$.} Wilson loops (WLs) are also available to quantitatively differentiate non-trivial orders \cite{Wang2010,Tchernyshyov2006}. These gauge-invariant quantities are defined along a given closed path on the lattice. Here, two types of non-winding loops are required to categorise the Ans\"atze by their flux structure: loops of length 6 on a hexagon, and of length 8 on a rhombus [Fig.~\ref{fig:2}]. Magnetic phases are now characterised by the flux piercing each of these loops $(\phi_h / \pi , \phi_r/ \pi)$ \cite{schaffer2017}, with $\phi_h = \arg (A_{12}(-A_{25}^*)A_{57}(-A_{76}^*)A_{64}(-A_{41}^*) )$ and $\phi_r = \arg (A_{01}(-A_{12}^*)A_{25}(-A_{58}^*)A_{87}(-A_{76}^*)A_{64}(-A_{40}^*) )$.\\

\textbf{The phase diagram} obtained from SBMFT is composed of three phases [Fig.~\ref{fig:2}] and is consistent with the pf-FRG results of \cite{buessen2016}, with two spin liquids and an ordered phase. Ref.~\cite{buessen2016} was, however, addressing a large range of models, and the nature of the spin liquids and position of the boundaries were not necessarily discussed in details. Also, while our Schwinger-boson approach is a zero-temperature mean-field theory, pf-FRG results were obtained at low, but nonetheless finite, temperatures. The structure factors of Fig.~20 in \cite{buessen2016} are for example reminiscent of Fig.~3 in \cite{mizoguchi2018} obtained from classical Monte Carlo simulations at low temperature. A precise comparison between our two works is thus difficult. With that in mind, the SBMFT phase diagram is:

\textit{Chiral spin liquid cuboc1}:
At $J=0$ the HKA ground state within SBMFT is known to be the \textit{cuboc1} state \cite{messio2012}. The name comes from its magnetic unit cell composed of 12 spins forming the shape of a cuboctahedron. \red{Since the flux piercing an hexagon is not quantized in units of $\pi$, the phase is chiral and breaks the time-reversal symmetry. We find that the chiral \textit{cuboc1} ground state persists up to $J=0.33$, whose Ansatz possesses a gauge degree of freedom \cite{SupMat}.}

\textit{$Z_{2}^{(1,0)}$ spin liquid}: 
At $J=0.33(1)$, a phase transition to a $Z_2$ QSL is observed. This phase has the same flux structure $(\pi,0)$ than the gapped SL obtained from the quantum melting of the $q=0$ order introduced by \cite{Sachdev1992}. All mean field parameters have the same amplitudes $\mathcal{A}$ and $\mathcal{B}$, whose values slowly vary with $0.33 < J < 0.50$ while preserving the $(\pi,0)$ flux structure.

\textit{Chiral magnetic order LRO$^{(1,1)}$}:
A second transition takes place at $J=0.50(1)$, concomitant with the closing of the gap $\Delta$ in the thermodynamic limit [Fig.~\ref{fig:2}.(d)], indicating long-range order, with both hexagons and rhombii WLs possessing a $\pi$ flux. This state can be seen as a Bose condensation of the $Z_2^{(1,1)}$ QSL reported by \cite{schaffer2017} in the breathing kagome lattice; we call it the LRO$^{(1,1)}$ state. Additionally, $\arg(\mathcal{A})$ and $\arg(\mathcal{B})$ are non zero, which means that this magnetic order is chiral.\\

\begin{figure}[t]
\centering\includegraphics[width=0.5\textwidth]{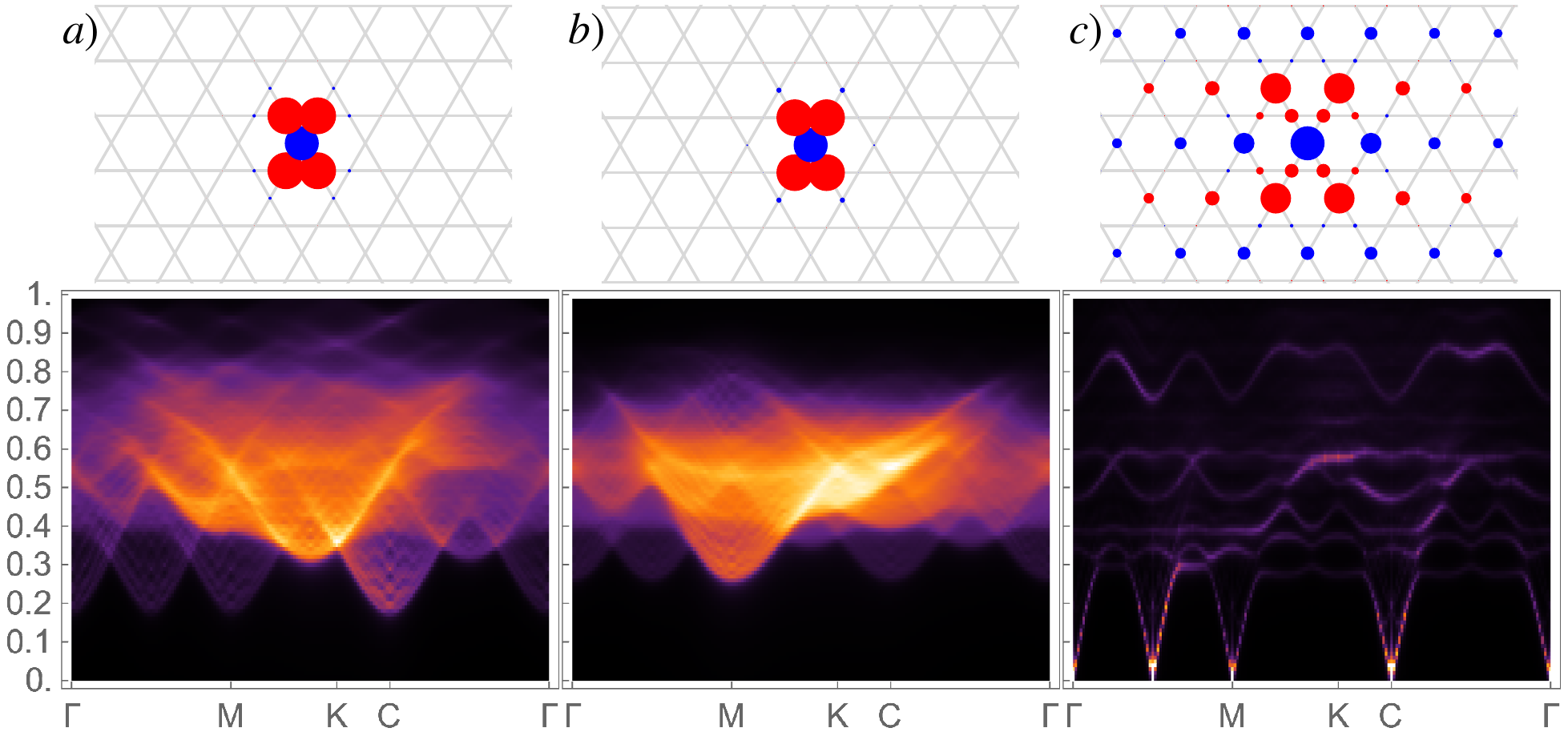}
\caption{
\red{Correlations in real space (top) and Fourier space $S(\mathbf{q},\omega)$ [Eq.~(\ref{eq:Sqw})] (bottom) for the (a) \textit{cuboc1} at $J=0.2$, (b) $Z_{2}^{(1,0)}$ at $J=0.4$ and (c) LRO$^{(1,1)}$ at $J=0.6$.} The reference site is the blue circle at the centre. The strength and sign of correlations are respectively given by the radius and colour of the circles (red is negative). \red{The path in Fourier space is given in Fig.~\ref{fig:1}.} System size is $n_{s}=1728$.
}
\label{fig:corr}
\end{figure}

\begin{figure*}[ht]
\centering\includegraphics[width=0.9\textwidth]{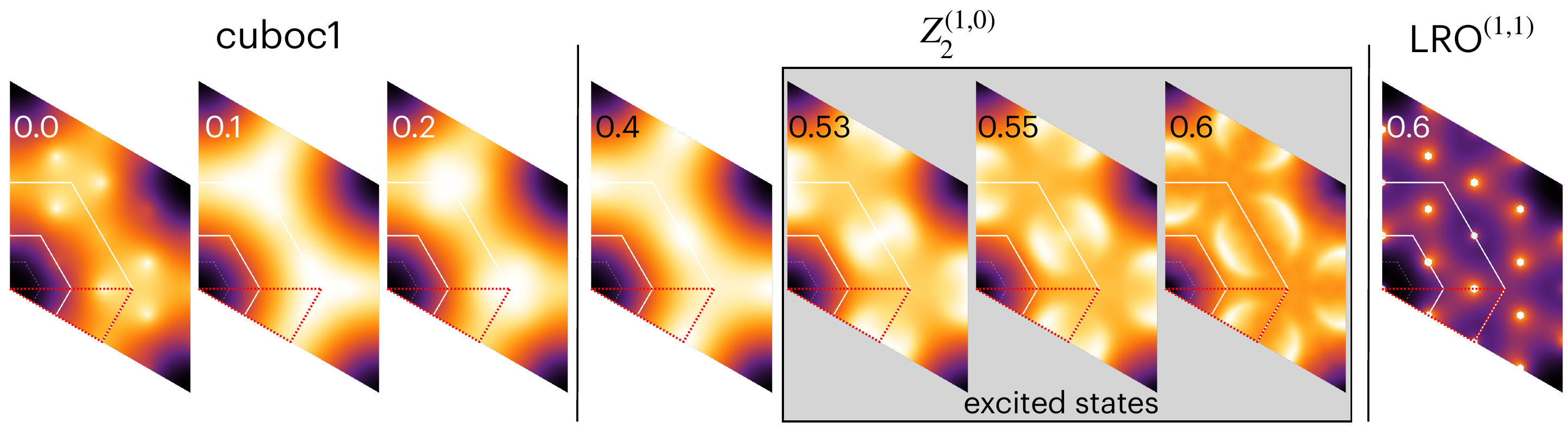}
\caption{\red{
Structure factor  $S(\mathbf{q})$ of the \textit{cuboc1}, $Z_2^{(1,0)}$ and LRO$^{(1,1)}$ Ans\"atze for increasing $J$ (given in each panel). The $Z_{2}^{(1,0)}$ structure factor is qualitatively the same for $0.33<J \leq 0.5$, but when the $Z_{2}^{(1,0)}$ Ansatz becomes an excitation for $J>0.50$ (grey boxed panels), half-moon patterns appear in the 2nd BZs. The color scale is normalised on all panels, except for LRO$^{(1,1)}$ where a cutoff is imposed to emphasize the low-intensity scattering. System size is $n_{s}=1728$.
}}
\label{fig:SQ}
\end{figure*}

\textbf{Discussion.}
In the rest of this paper we will endeavour to rationalise the origin of these phase transitions. Let us start with the onset of magnetic order at $J=0.5$. Up to a constant, Eq.~(\ref{eq:ham}) is equivalent to \cite{Ishizuka13c,mizoguchi2017,mizoguchi2018}
\begin{eqnarray}
\mathcal{H}=\left(\frac{1}{2}-J\right)\sum_{n} \hat\mathbf{M}_{n}^{2}\, -\,
J\sum_{\langle n,m\rangle}\hat\mathbf{M}_{n}\cdot\hat\mathbf{M}_{m}
\label{eq:hamcl}
\end{eqnarray}
where the summations run over all triangles $n$ and neighbouring pairs of triangles $\langle n,m\rangle$, and
\begin{eqnarray}
\hat\mathbf{M}_{n}\equiv \zeta_{n}\sum_{i\in n}\hat\mathbf{S}_{i}
\label{eq:Mn}
\end{eqnarray}
is the magnetisation of the three spins of triangle $n$, up to a staggered prefactor $\zeta_{n}=\pm 1$ distinguishing between up and down triangles. This mapping was studied classically on the kagome \cite{mizoguchi2017,mizoguchi2018} and pyrochlore \cite{rau16b,udagawa2016} lattices. For $J>1/2$, the first term of Eq.~(\ref{eq:hamcl}) favours saturated magnetisation on all triangles, while the second term prevents long-range ferromagnetism because of the staggered prefactor $\zeta_{n}$. For Ising spins, $\hat\mathbf{M}_{n}$ becomes a discretised scalar corresponding to a topological charge sitting on all triangles, and the first term of Eq.~(\ref{eq:hamcl}) is their chemical potential. Hence, at the level of a triangle, $J=1/2$ is the frontier between locally antiferromagnetic ($J<1/2$) and ferromagnetic ($J>1/2$) fluctuations. This interpretation is in agreement with real-space correlations [Fig.~\ref{fig:corr}]. Nearest-neighbour correlations are short-range antiferromagnetic in the \red{\textit{cuboc1} and} $Z_{2}^{(1,0)}$ spin liquid [Fig.~\ref{fig:corr}.(a,b)], while ferromagnetic correlations appear on some triangles in the LRO$^{(1,1)}$ Ansatz [Fig.~\ref{fig:corr}.c]. The position of this boundary could a priori be shifted by the second term of Eq.~(\ref{eq:hamcl}) -- this is what happens in classical systems \cite{mizoguchi2017,mizoguchi2018} -- but in our quantum model, this local mechanism is a probable cause for the Bose condensation observed at $J=0.5$.

\red{When taken alone, the $J_{3}$ term connects only spins on the same sublattice and forms three disconnected \textit{non-frustrated} square lattices, which explains the long-range antiferromagnetic order between same-sublattice spins of Fig.~\ref{fig:corr}.c. And the loss of correlations between different sublattices at long distance indicates destructive quantum interference, probably due to the $J_{2}$ term connecting different sublattices, that forms three disconnected \textit{frustrated} kagome lattices.
}

In classical systems at low temperature, this onset of local ferromagnetism coincides with the apparition of characteristic patterns in the structure factor, known as half moons \cite{mizoguchi2018}, that were also observed in pf-FRG \cite{buessen2016}. Here we do not find these patterns in the ground-state phase diagram [Fig.~\ref{fig:SQ}]. However, for $J>0.5$, the lowest excited Ansatz we could stabilise in the self-consistent SBMFT procedure is the $Z_{2}^{(1,0)}$ spin liquid, with an additional chiral flavour (i.e. some of its mean-field parameters become complex for $J>0.5$). This chirality coexists with half-moon patterns in the structure factor. One needs to remain cautious since SBMFT is a zero-temperature calculation, but the presence of these patterns in an excited Ansatz is consistent with their presence at low temperature \cite{mizoguchi2018,buessen2016}.\\

On the other end of the phase diagram, the presence of the \textit{cuboc1} phase corresponds to the region of stability of the HKA spin liquid, within SBMFT, in presence of the $J$ perturbation. This region is noticeably large \cite{buessen2016,li2021} and raises the question about the origin of such permanence. Let us consider another rewriting of Hamiltonian (\ref{eq:ham}), up to a constant
\begin{eqnarray}
\mathcal{H}=\frac{1}{2}\left(1-3J\right)\sum_{n} \hat\mathbf{T}_{n}^{2}\, +\,
\frac{J}{2}\sum_{\langle \alpha \rangle}\hat\mathbf{G}_{\alpha}^{2}
\label{eq:hamhg}
\end{eqnarray}
where the summations run over all triangles $n$ and bi-triangles $\alpha$. A bi-triangle is composed of two triangles and 5 sites (in a shape reminiscent of a hourglass).
\begin{eqnarray}
\hat\mathbf{T}_{n}\equiv \sum_{i\in n}\hat\mathbf{S}_{i}
\qquad\mathrm{and}\qquad
\hat\mathbf{G}_{\alpha}\equiv \sum_{j\in \alpha}\hat\mathbf{S}_{j}
\label{eq:Mn}
\end{eqnarray}
are the total magnetisation on triangle $n$ and bi-triangle $\alpha$. According to the fusion rule of angular momentum, we have $\langle \hat\mathbf{T}_{n}^{2}\rangle=T_{n}(T_{n}+1)\in\{\frac{3}{4},\frac{15}{4}\}$ and $\langle \hat\mathbf{G}_{\alpha}^{2}\rangle=G_{\alpha}(G_{\alpha}+1)\in\{\frac{3}{4},\frac{15}{4},\frac{35}{4}\}$. Hence, for $0< J<1/3$, the minimal eigenvalue of Hamiltonian (\ref{eq:hamhg}) would be, if geometrically possible,
\begin{eqnarray}
\{T_{n}=\frac{1}{2}\;\&\; G_{\alpha}=\frac{1}{2}\,|\,\forall\, n,\alpha\}.
\label{eq:gs}
\end{eqnarray}
Eq.~(\ref{eq:gs}) means having one singlet on all triangles and two singlets on all bi-triangles, with the important property that the former constraint is a sufficient condition to satisfy the latter. Paving the kagome lattice with one singlet per triangle is famously impossible; otherwise the HKA ground state for $J=0$ would have been known for a long time. That being said, we know that the HKA ground state, irrespectively of its nature, necessarily minimises the energy of the first term of Hamiltonian (\ref{eq:hamhg}). According to constraint~(\ref{eq:gs}), we can reasonably expect that the HKA ground state would also minimise the second term of Hamiltonian (\ref{eq:hamhg}), up to a small deformation of the Ansatz. And in that case, this deformed HKA Ansatz should remain ground state up to $J\sim 1/3$ when constraint~(\ref{eq:gs}) stops to be valid. This is precisely what we find in SBMFT with \red{a stable \textit{cuboc1} phase and a slowly varying chiral $\phi_{h}$ flux from $J=0$ to $0.33$ (see \cite{SupMat})}. Please note that the almost perfect match is probably too good to be true; corrections beyond mean field might shift the boundary a little bit. Fermionic mean-field theory finds e.g.~a boundary at $J=0.27$ \cite{li2021}, where the stability of the $J=0$ phase is related to the persistence of the eigenbasis of the mean-field Hamiltonian. \red{For the sake of completeness, we checked that the phase diagram remains essentially the same for larger spins $S=1/2$, with boundaries shifted from 0.33 to 0.28 and from 0.50 to 0.47 \cite{SupMat}. The main difference is that the HKA ground state is known to be gapless for $S=1/2$ within SBMFT \cite{messio2012}. In our model, the gap opens at $J=0.15-0.20$.}\\

\textbf{Summary} We find that (i) the HKA spin liquid persist up to $J\approx 1/3$ because the further-neighbour perturbation $J$ is only weakly frustrated with the dominant nearest-neighbour antiferromagnetic exchange; and (ii) the onset of long-range order at $J\approx 1/2$ comes from the  local change of fluctuations imposed on each triangle. In that sense, the intermediate $Z_{2}^{(1,0)}$ spin liquid is the best compromise within SBMFT satisfying the competition between the two terms of Hamiltonians (\ref{eq:hamcl}) and (\ref{eq:hamhg}). As a side note, the $Z_{2}^{(1,0)}$ spin liquid also coincides with a shift of scattering from the K to the M points in the structure factor, a precursor of the emergence of the half moons in the excited Ansatz for $J>1/2$ [Fig.~\ref{fig:SQ}].\\

\textbf{Outlook} \red{The HKA ground state is famously elusive \cite{yan11a,Iqbal11a,messio2012,Mendels16,Lauchli16a,He17a,Ralko2018}. While magnetic order is likely stable above $J\approx 1/2$ beyond mean field, the low$-J$ phase is expected to depend on the method. In this context, does our reasoning based on Hamiltonian (\ref{eq:hamhg}) remain valid with other methods, and the HKA spin liquid persisting up to $J\sim 1/3$ ? The dynamics, excitations and low-temperature physics are also promising directions to follow.}

To conclude, we should point out that constraint (\ref{eq:gs}) is easily satisfied on a Husimi cactus. It means that the rewriting of Hamiltonian (\ref{eq:hamhg}) provides an exact solution of the ground state of a non-trivial interacting model on a Husimi cactus for $0\leq J\leq 1/3$, that can be extended to other geometries. The line $J_{2}=J_{3}$ probably shares some common properties across different frustrated systems in various dimensions that would be worth exploring in a systematic way.\\


\textbf{Acknowledgments} 
We acknowledge Yasir Iqbal, Tao Li, Laura Messio, Tommaso Roscilde, Julia Meyer and Sylvain Capponi for useful discussions. L.J. acknowledges financial support from CNRS (PICS No. 228338) and from the French ``Agence Nationale de la Recherche'' under Grant No. ANR-18-CE30-0011-01, and A.R. under Grant No. ANR-21-CE30-0029-01. M. U. is supported by JSPS KAKENHI (Nos. 20H05655 and 22H01147), MEXT, Japan.\\

\red{
\textit{Note added after submission:} When submitting this manuscript, a preprint by Kiese \textit{et al} \cite{Kiese22} appeared studying the same model with complementary methods. They also find long-range magnetic order for $J>0.51(1)$, which is consistent with our real-space correlations [Fig.~\ref{fig:corr}.c] and supports the same Bragg peaks (except at the M point) and low-intensity star patterns as in Fig.~\ref{fig:SQ} for LRO$^{(1,1)}$. The HKA ground state obtained from their methods is the U(1) gapless spin liquid, which persists up to $J=0.30(5)$, consistent with our estimate of $\sim 1/3$. Their intermediate phase, however, is not a spin liquid but a pinwheel valence bond crystal.
}

\bibliographystyle{apsrev4-1} 
\bibliography{references} 

\end{document}